\documentclass[conference]{IEEEtran}
\IEEEoverridecommandlockouts

\usepackage{import}

\usepackage{cite}
\usepackage{amsmath,amssymb,amsfonts}
\usepackage{graphicx}
\usepackage{textcomp}
\usepackage{xcolor}

\usepackage{macros}
\usepackage{amsthm}

\usepackage{caption}
\usepackage{subcaption}
\usepackage{hyperref}
\usepackage{booktabs}
\usepackage{array, multirow}
\usepackage[flushleft]{threeparttable}
\usepackage{tablefootnote}

\usepackage{bm}
\usepackage{mathrsfs}

\newtheoremstyle{colon}%
{}
{}
{\itshape}
{}
{\bfseries}
{:}
{ }
{}

\theoremstyle{colon}
\newtheorem*{casestudy}{Case Study}

\usepackage[export]{adjustbox}

\usepackage{flushend}
\begin{document}

\title{Trust in Motion: Capturing Trust Ascendancy in Open-Source Projects using Hybrid AI
\thanks{This material is based upon work supported by the Defense Advanced Research
  Projects Agency (DARPA) under Contract No.~HR00112190086. Any opinions,
  findings, and conclusions or recommendations expressed in this material are those of
  the author(s) and do not necessarily reflect the views of the United States Government or DARPA.}
}

 \author{
   \IEEEauthorblockN{Huascar Sanchez}
   \IEEEauthorblockA{
     \textit{Computer Science Laboratory}\\
     \textit{SRI International}\\
     huascar.sanchez@sri.com
   }
 \and
   \IEEEauthorblockN{Briland Hitaj}
   \IEEEauthorblockA{
     \textit{Computer Science Laboratory} \\
     \textit{SRI International}\\
     briland.hitaj@sri.com
   }
 }

\maketitle

\thispagestyle{plain}
\pagestyle{plain}

\begin{abstract}

Open-source is frequently described as a driver for unprecedented communication and
collaboration, and the process works best when projects support teamwork. Yet,
open-source cooperation processes in no way protect project contributors from
considerations of trust, power, and influence. Indeed, achieving the level of trust
necessary to contribute to a project and thus influence its direction is a constant
process of change, and developers take many different routes over many communication
channels to achieve it. We refer to this process of influence-seeking and trust-building
as \textit{trust ascendancy}.

This paper describes a methodology for understanding the notion of trust ascendancy and
introduces the capabilities that are needed to localize trust ascendancy operations
happening over open-source projects. Much of the prior work in understanding trust in
open-source software development has focused on a static view of the problem using
different forms of quantity measures.  However, trust ascendancy is not static, but
rather adapts to changes in the open-source ecosystem in response to new input. This
paper is the first attempt to articulate and study these signals from a dynamic view of
the problem. In that respect, we identify related work that may help illuminate research
challenges, implementation tradeoffs, and complementary solutions. Our preliminary
results show the effectiveness of our method at capturing the trust ascendancy developed
by individuals involved in a well-documented 2020 social engineering attack. Our future
plans highlight research challenges and encourage cross-disciplinary collaboration to
create more automated, accurate, and efficient ways to model and then track trust
ascendancy in open-source projects.

\end{abstract}


\section{Introduction}%
\label{sec:intro}

Achieving the level of trust necessary to contribute to a project is a ubiquitous
construct of how open-source software development works~\cite{scacchi2007free,
von2003community} and one of the most prevalent objectives~\cite{nagle2020report} in the
general developer population in social coding platforms like GitHub and Stack Overflow.
Achieving this trust is a dynamic process of change~\cite{de2010can} that is inherently
political~\cite{ducheneaut2005socialization}, and developers take many different routes
over many communication channels to influence its formation~\cite{von2003community,
dabbish2012social, jiang2014tracing, bettenburg2015management, wang2016diffusion}. We
refer to this process of influence-seeking and trust-building as \textit{trust
ascendancy}. Much of the prior work in understanding trust and its ascendancy in
open-source projects has focused on a static view of the problem using scale
measurements (e.g.,~\cite{gysin2010trustability, sinha2011entering,
calefato2017preliminary, wermke2022committed}). However, trust ascendancy is not static.
Instead, it adapts to changes in the ecosystem in response to developer role changes,
new functionality, new technologies, and so on. Automatically tracking this
socio-technically stimulated dynamism thus requires dynamic developer behavior models.
This paper is a first attempt to articulate and study this issue.

We consider the problem of capturing the \textit{motion dynamics} of trust ascendancy
inside open-source software (OSS) projects using dynamic developer activity models.
These motion dynamics are reflected in the way trust is periodically developed inside
projects in response to either socio-technical stimuli (e.g., social influence, role
changes, code contributions) or to periodic changes in the context of individual
activities, such as reporting a bug, that are intended to help potential contributors
build a reputation and eventually become project committers (see the case study of a
``successful socialization'' in Ducheneaut~\cite{ducheneaut2005socialization}).
Understanding the context in which actions are performed as well as tracking
\textit{when} (e.g., time of day) this context changes can give us a global picture of
the \textit{influence pathways} formed inside a project. Here, an influence pathway is a
potential conduit for influence to flow inside a project and a schedule. The context of
an activity embodies the \textit{semantic associations} between the activity and other
activities that were performed around the same schedule.

Arguably, influence is the main driver for building trust inside \textit{networked
social environments}~\cite{wang2016diffusion} like OSS projects. The structure of these
networks is usually ``black-boxed,'' and to exert any influence in them, potential
contributors need to progressively make this network structure more
visible~\cite{ducheneaut2005socialization}. A goal of the current effort is to bridge
the gap mentioned earlier, namely, that research on understanding trust and its
ascendancy tends to be based on static accounts. Consistent with this goal, we introduce
a hybrid approach that combines the strength of unsupervised machine learning with the
flexibility of self-supervised machine learning, and generalize it to sequential data
collected from real-world software projects. This work complements existing work by
providing better mapping and understanding of the multiple influence pathways taken by
developers to progressively open this ``black-box,'' thereby enabling them to
contribute to the project.

\begin{figure}[t!]
  \centering
    \begin{subfigure}{\columnwidth}
      \centering
      \includegraphics[height=.7\columnwidth, frame]{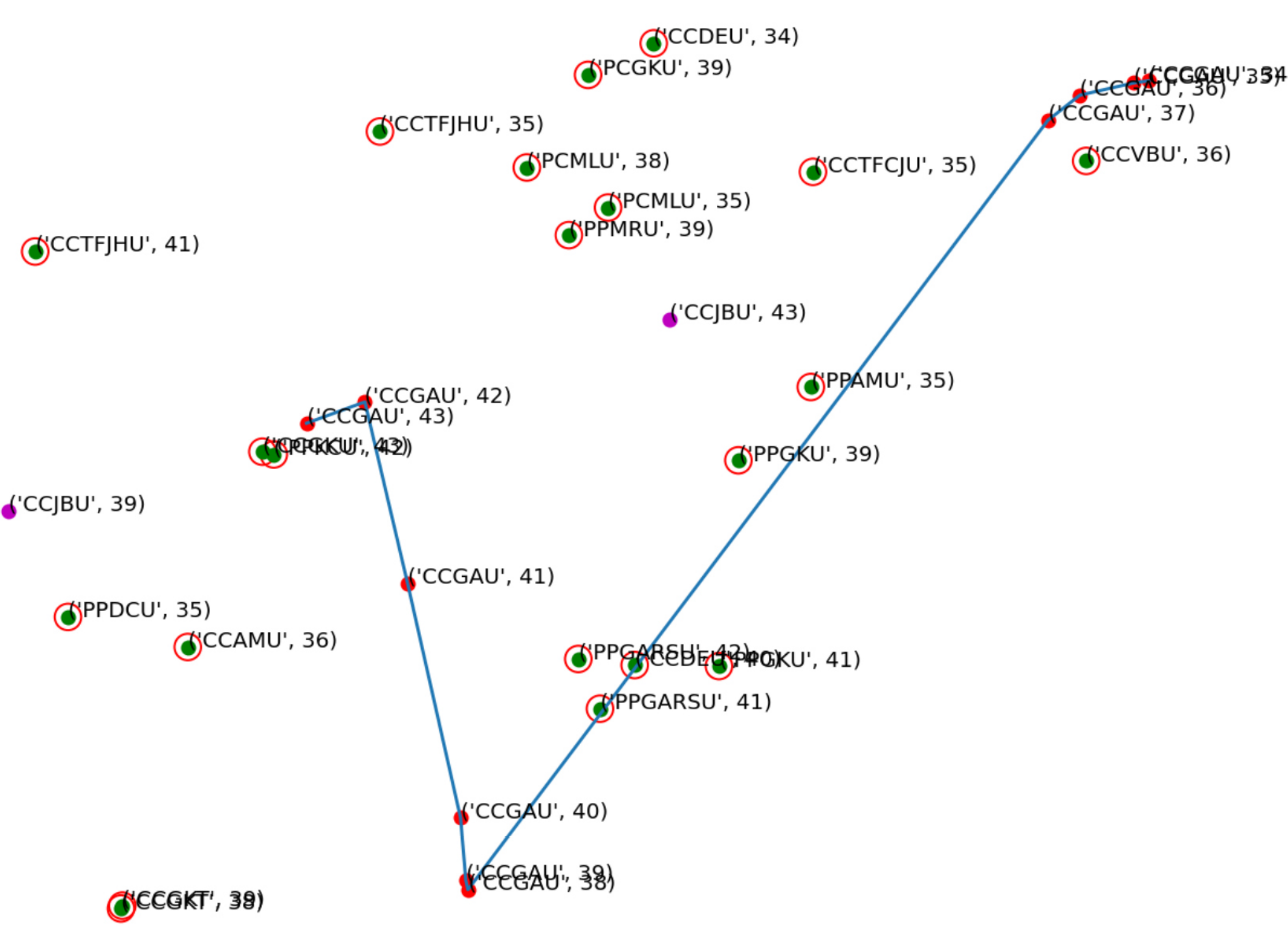}
      \caption{Opportunistic trust ascendancy.}%
      \label{fig:acosta}
    \end{subfigure}
    \begin{subfigure}{\columnwidth}
      \centering
      \includegraphics[height=.7\columnwidth, frame, scale=0.82]{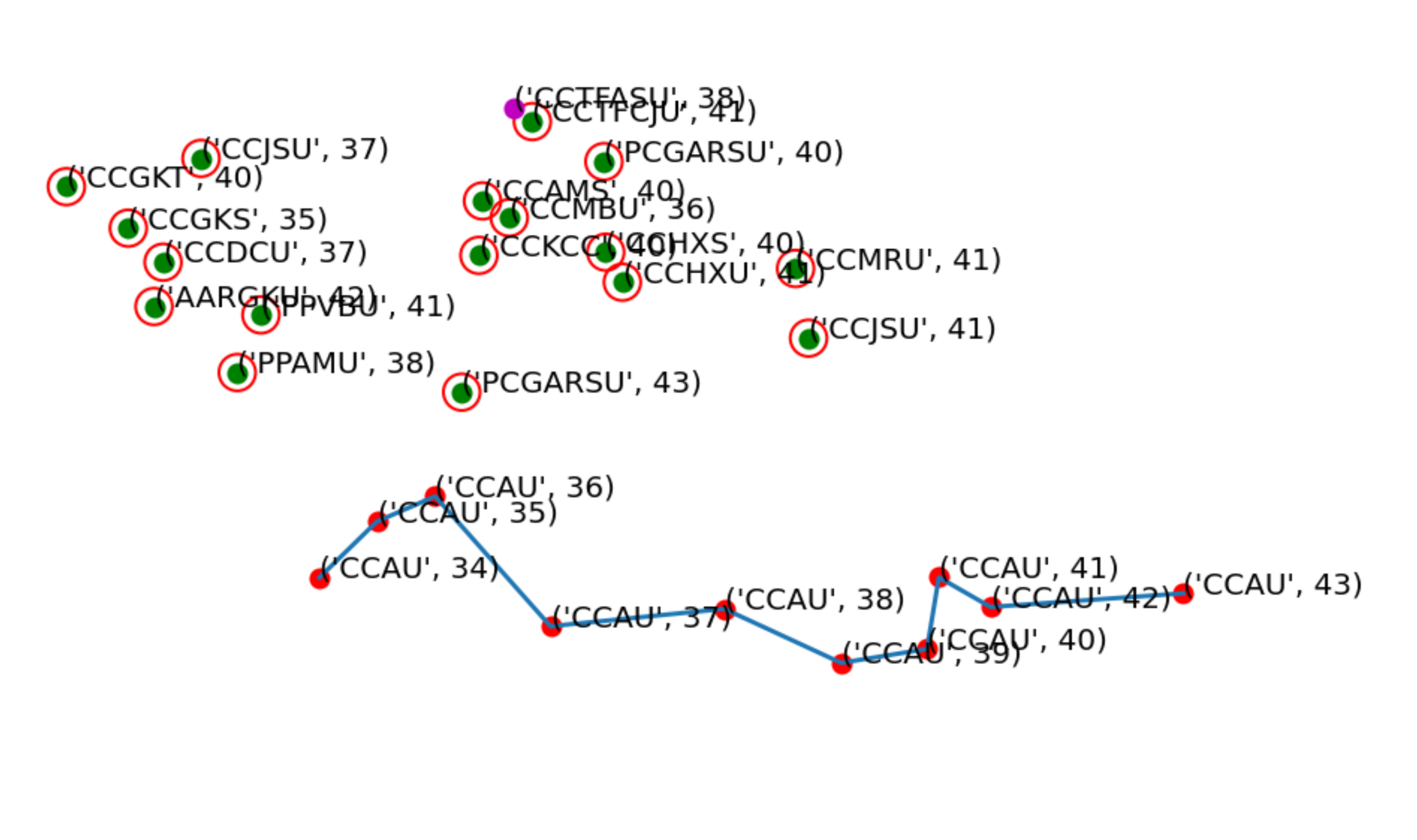}
      \caption{Awry trust ascendancy.}%
      \label{fig:aditya}
    \end{subfigure}
    \begin{subfigure}{\columnwidth}
      \centering
      \includegraphics[height=.7\columnwidth, frame, scale=0.85]{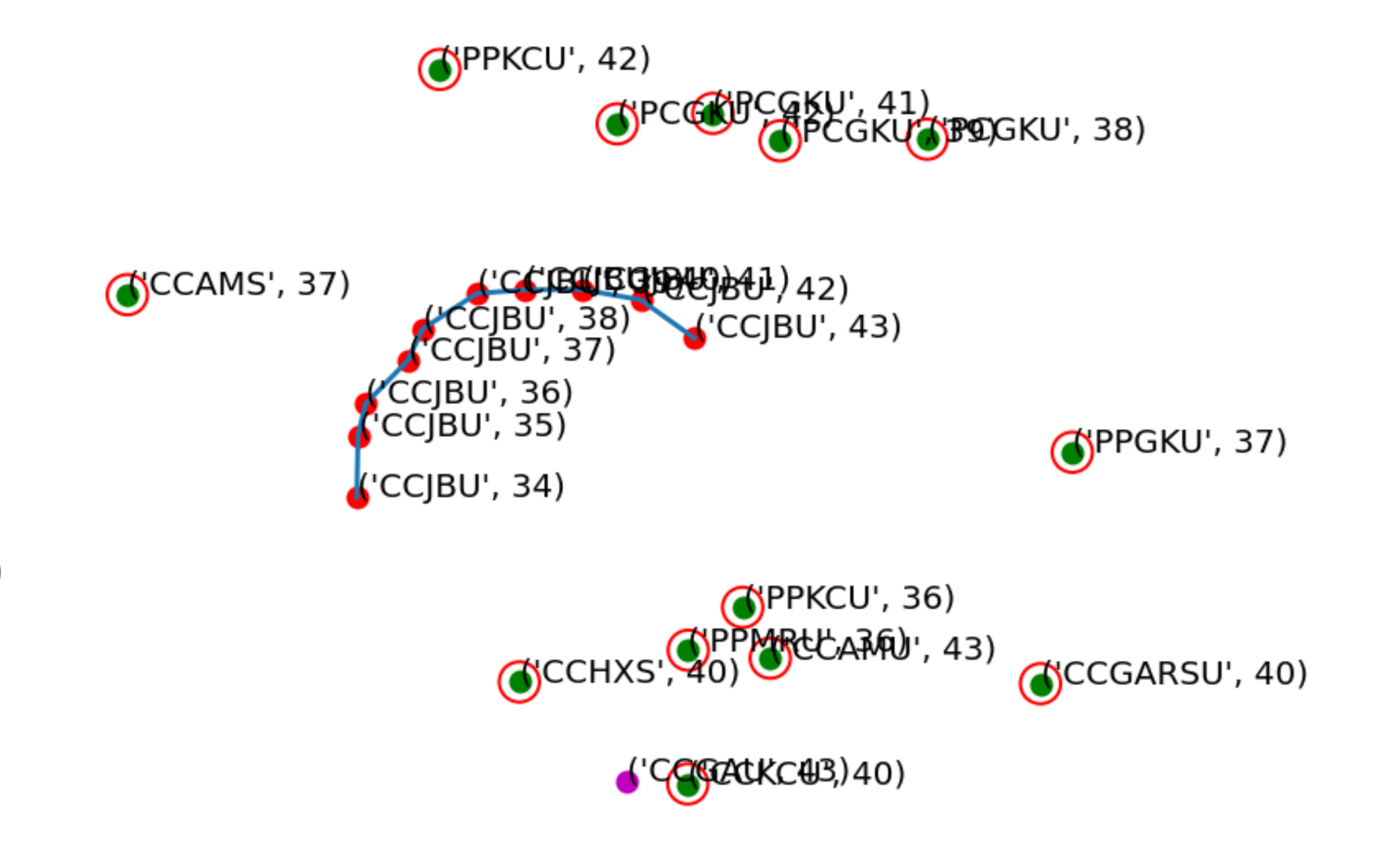}
      \caption{Hit or miss trust ascendancy.}%
      \label{fig:bonds}
    \end{subfigure}
    \caption{A $2$D t-SNE projection of three trust ascendancy trajectories across a
      $14$-week time window ($08/2020-11/2020$). Red dots represent attackers' behavior,
      green dots with a red ring represent maintainers' behavior, purple dots represent
      the behavior of other aliases used by the attackers, labels are action
      abbreviations, and blue lines represent the trajectory of attackers' trust
      ascendancy. Maintainers' distance to these trajectories is an indicator of trust
      development with attackers' actions.}%
    \label{fig:three trajectories}
\end{figure}

When we look at existing work in language evolution detection using word
embeddings~\cite{kulkarni2015statistically, bamler2017dynamic, yao2018dynamic,
rudolph2018dynamic, di2019training, hofmann-etal-2021-dynamic}, we observe the power of
word embeddings at capturing evolving semantic associations between words across time
and also at allowing for cross-pollination between the NLP and other application domains
(e.g.,~\cite{mahdavi2018dynnode2vec, shenattack2vec2019, chen2019dynamic,
csuvik2019source, dakhel2022dev2vec}). In general, these models accurately model the
distribution of an object (e.g., words, security alerts) based on the surrounding
objects in terms of low-dimensional vector representations. In building on this prior
work, we are interested in learning dynamic developer activity embeddings for evolving
influence pathways recovery. These embeddings will capture, in a fashion that resembles
how dynamic word embeddings infer embedding trajectories over time, the temporal
dependence between concrete developer actions performed within OSS projects. This will
give us a natural \textit{backtrace trajectory} of the influence pathways taken by
potential contributors. Figure~\ref{fig:three trajectories} shows a concrete
example of how we can use the dynamic developer activity embeddings to reveal distinctly
novel influence trajectories that summarize a $2020$ social engineering attack against
the Linux Kernel~\cite{lkml_Cook_2021}.

The revealed trajectories show the potentially \textit{influenced}~\footnote{Manipulated
directly or indirectly in response to other maintainers' actions.}  maintainers and the
attackers' trust ascendancy operations across both time and subsystems. The objective of each
operation was to build trust and then to poison the Linux Kernel with vulnerabilities
(the ``hypocrite commits'' patches~\cite{wu2021feasibility}). The sections that follow
provide details of the case study we explored to highlight the challenge, as well as the
methods and data we used to recover these dynamic trajectories.

\section{Case study: Highlighting Characteristics of the Challenge}%
\label{subsec:casestudy}

The techniques described in this paper should apply to thousands of significant OSS
projects. We considered many projects, but narrowed the list of case studies to
\textit{the Linux Kernel and its accepted patches} case study below.

\begin{casestudy}

\textit{The Linux Kernel and its accepted patches.} The Linux Kernel (LK) development
process is well known, documented, and researched~\cite{kroahhartman2007linux,
filippova2016effects}. However, some actions of contributors or maintainers often
diverge from this process. One of these actions is accepting patches. Studying why a
technical change happens is as important as modeling either the rejection or the
oversight of design changes (e.g.,~\cite{riehle2019analysis}), particularly if the
acceptance is rooted in both social and technical influence. To provide a control
comparison, we will model the influence pathways that have led to the integration of
controversial changes such as the \textit{hypocrite commits}~\cite{wu2021feasibility}.
\textit{Could this event be indicative of an intrinsic susceptibility to socio-technical
influence in general OSS projects?}

\end{casestudy}

We rely on the high level of transparency offered by the kernel and the wealth of
signals available in its development process to learn a dynamic model of developer
behavior.

\section{Methodology}%
\label{sec:method}

In this section, we describe our methodology for answering the question of trust
ascendancy in OSS projects.

The \textbf{basics of our methodology} are thus: We consider a large OSS project as a
case study to highlight the characteristics of the challenge. This case study project
experienced an important code change that was associated with a form of social
influence, and each case was well documented and widely discussed. Because of the
heterogeneity of the types of contributors (e.g., domain experts, hackers, superficial
contributors, vandals) who can participate in a large project, we expect the developer
data to be scattered or noisy. To address this issue, we formulate our exploration task
as a process of identifying a subset of general types of developer activity that best
explain the data. In that regard, we use unsupervised machine learning to extract
general activity types and then generalize them to sequential data. Having done that, we
use self-supervised machine learning to generate properly aligned temporal embeddings of
developer activities.

\subsection{Activity-based Analysis of Developer Activity Data}%
\label{subsec:activities}

To capture the motion dynamics of trust ascendancy in the kernel, we need to understand
first its activity space and any inter-relations that hold among them. To this end, we
consider a snapshot of its development history, pulled directly from the Linux Kernel
mailing list~\footnote{\url{https://lkml.org/}} (LKML). This snapshot contains
timestamped information pertaining to concrete developer actions (e.g., submitting a
patch, acknowledging a submitted patch) in the LKML, in a $2019-2021$ time window.
Furthermore, the collected patch emails should have at least one email reply in the LKML
and at least $50$ words (or at least one sentence) in their email body, persuade its
recipients to accept them~\footnote{Detected persuasion in emails using a transfer
learning setup for Random Forest~\cite{segev2016learn}, trained on the Persuasion4Good
dataset~\cite{wang2019persuasion} and LKML data.}, and be sent by a human (i.e., a non
bot).

\begin{figure}[th!]
  \centering
  \includegraphics[width=1\columnwidth]{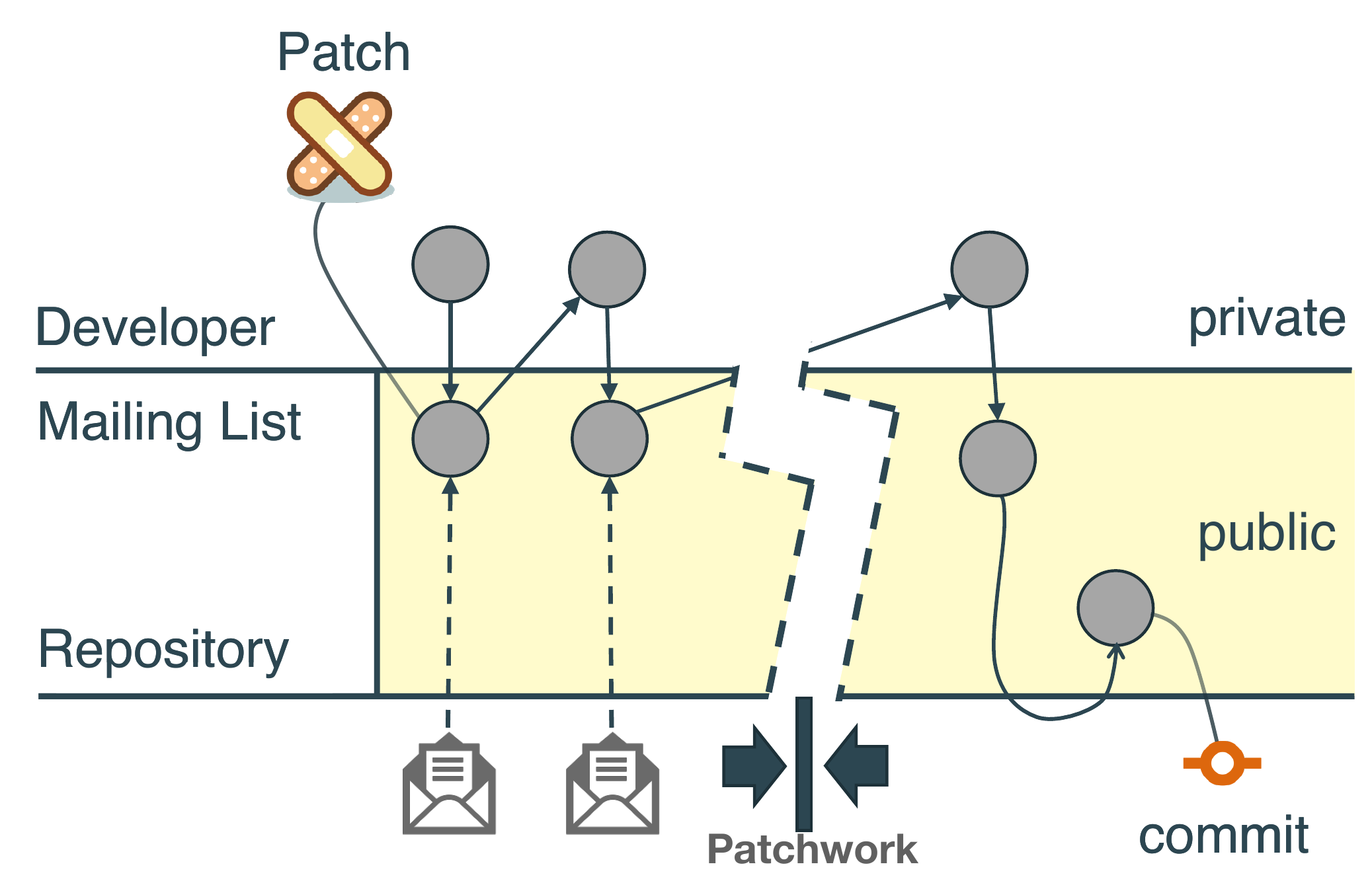}
  \caption{Bridging the gap between LKML and LK's codebase via Patchwork
    (adapted from image in~\cite{riehle2019analysis}).}%
  \label{fig:patchwork}
\end{figure}

As illustrated in Figure~\ref{fig:patchwork}, this link between patch emails in LKML and
commits in LK is nonexistent and thus requires an indirect provider. We circumvent this
problem by probing the \textit{patchwork}
service~\footnote{\url{https://patchwork.kernel.org/}} according to the process
described in Xu and Zhou's paper~\cite{xu2018multi}. Overall, we collected a total of
$27411$ records matching our patch email criteria.

We assess the collected data using different sets of characteristics, each covering a
(mail-based) software development aspect: (1) contribution, (2) exposition, and (3)
administration; see Table~\ref{tab:metrics} for a summary. These aspects are
interrelated, and thus their enclosed characteristics can be assembled into descriptive
action categories using Exploratory Factor Analysis~\cite{child2006essentials} (EFA)
according to the process described in Cheng and Guo~\cite{cheng2019activity}. In EFA
language, these descriptive action categories are known as \textit{latent factors}.

\begin{table*}
  \centering
  \caption{Characteristics describing developer actions.}%
  \label{tab:metrics}
  \begin{threeparttable}[b]
    \centering
    \begin{tabular}{rl>{\raggedright\arraybackslash}p{5cm}}
      \hline
      \textbf{Type} & \textbf{Characteristic} & \textbf{Description} \\ \hline
      Contribution & \textit{Sender Experience} & \#accepted patches / \#submitted patches \\
      & \textit{Sender Engagement} & (\#new threads -\@ \#bot spam) / \#sent emails \\
      & \textit{Persuasive} & Is exerting influence via persuasion? \\
      & \textit{Patch Email} & Is this email a patch email? \\
      & \textit{Bug Fix} & Is this a bug fix patch? \\
      & \textit{New Feature} & Is this a new feature patch? \\
      & \textit{Patch Churn} & Is this a new patch revision? \\ \hline
      Exposition & \textit{FKRE Score} & Text comprehension difficulty~\cite{schneider2016differentiating}\\
      & \textit{FKGL Score} & Text reading grade level~\cite{schneider2016differentiating}\\
      & \textit{Verbosity} & \#words / \#sentences \\ \hline
      Admin & \textit{Sent Time} & The time the patch email was sent \\
      & \textit{Received Time} & The time the patch email was received \\
      & \textit{First Patch Thread} & Is this the first patch email in email
                                        thread? \\
      & \textit{Accepted Patch} & Is this an accepted patch? \\
      & \textit{Accepted Commit} & Is this an integrated patch? \\ \hline
    \end{tabular}
  \end{threeparttable}
\end{table*}

Formally, we are given: $p$, the number of characteristics $(X_1, X_2, \ldots, X_p)$;
$m$, the number of latent factors $(F_1, F_2, \ldots, F_m)$; $X_j$, the representation
of each characteristic as a linear function of these factors plus a residual variance
(aka, latent errors) $e_j$, with \textit{the goal of producing a model that maximizes
the correlations between the characteristics (aka, observed variables) and the latent
factors in $X_j$}:

\begin{equation}\label{eq:efa}
\begin{gathered}
X_{j} = l_{j_{1}}\!F_1\;+\;l_{j_{2}}\!F_2\;+\;\ldots\;+\;l_{j_{m}}\!F_m\;+ e_j
\end{gathered}
\end{equation}

\noindent\textit{where $j = 1, 2, \ldots, p$}. The factor loadings
$l_{j_{1}}, l_{j_{2}}, \ldots, l_{j_{m}}$\footnote{$l_{j_{i}}$ is the correlation weight
of characteristic $j^{th}$ on the factor $1$.} in Equation~\ref{eq:efa} measure the
strength of the correlation between characteristics and
factors~\cite{child2006essentials}. The estimated values of the latent factors
$F_{i}, i > 0$ are known as factor scores.  By noting which characteristics are strongly
associated with each factor, we can interpret the meaning of the factors in $X_{j}$  and
use this knowledge to identify \textit{concrete action categories}.

\begin{table*}
  \centering
  \caption{Timestamped and labeled \textit{factor score vector} for Greg Kroah-Hartman (sender id $0$)}%
  \label{tab:factor-scaled-activity}
  \resizebox{.9\textwidth}{!}{%
  \begin{threeparttable}
  \begin{tabular}{cccccccc}
  \hline
    \textbf{sender\_id} & \textbf{sent\_time} & \textbf{Code Contribution}
    & \textbf{Knowledge Sharing} & \textbf{Patch Posting} & \textbf{Progress Control}
    & \textbf{Acknowledgment} & \textbf{label} \\ \hline
   0\tnote{$\star$} & \textbf{2020-08-20 09:35:52} & 0.83758650 & 0.00918697 &
   0.00502759 & 0.19685837 & 0.25811192 & $Y_{0}$ \\ \hline
  \end{tabular}%
  \end{threeparttable}
  }
\end{table*}

\begin{figure}[h!]
  \centering
  \includegraphics[width=1\columnwidth]{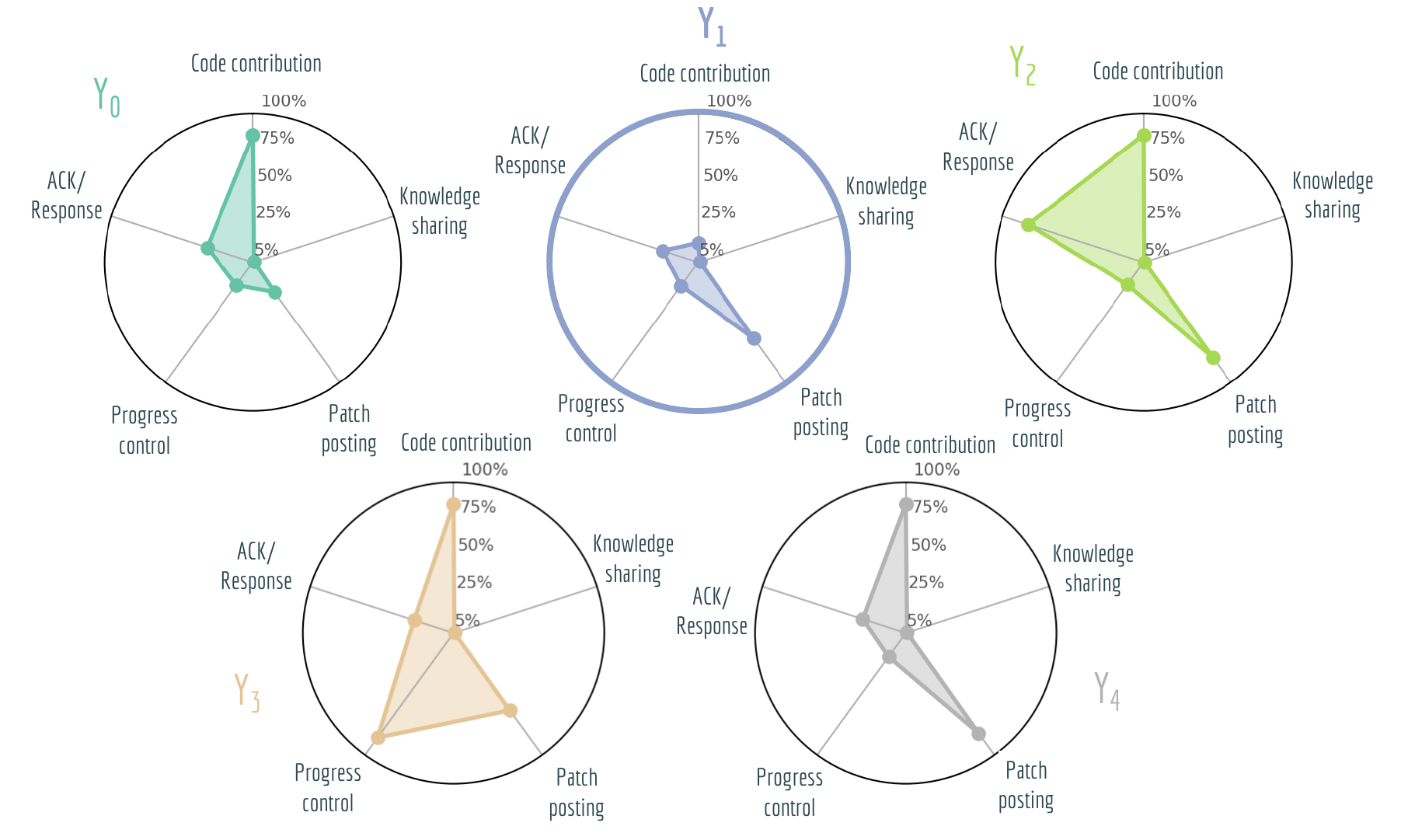}
  \caption{General types of developer activity, where $Y_{0}$ (contributing) describes a
    type of sporadic activities that occasionally lead to major contributions; $Y_{1}$
    (probing) describes a type of erratic activities that are followed by a flurry of
    subsequent emails; $Y_{2}$ (bug fixing) describes a type of periodic activities that
    consistently focus on small contributions like bug fixes; $Y_{3}$ (monitoring)
    describes a type of periodic activities that focus on checking progress; $Y_{4}$
    (persuading) describes a type of lone and often controversial activities that tend to
    instigate, persuade, or incite changes in project direction. We use these general
    activities as a compass for navigating and reading the trajectories revealed by our method.}%
  \label{fig:clusters}
\end{figure}

After applying EFA to our data, we interpreted $X_{j}$ to identify concrete action
categories. Doing so revealed five action types: \textit{(1) Code Contribution, (2)
Knowledge Sharing, (3) Patch Posting, (4) Progress Control, and (5) Acknowledgment /
Response}.  Like Cheng and Guo's work~\cite{cheng2019activity}, we use $X_{j}$'s factor
scores to extract clusters of action categories (see Figure~\ref{fig:clusters}). Without
loss of generality, we called these categories \textit{general developer activities}.
Unlike their work, we generalize the factor scores to \textit{sequential} data. Indeed,
we represent each timestamped event in the collected data as a \textit{row vector} of
non-zero factor scoring coefficients and is belonging cluster as its label.
Table~\ref{tab:factor-scaled-activity} provides an example of a timestamped developer
activity in August of $2020$.

\subsection{Dynamic Developer Activity Embeddings}%
\label{subsec:dynmodel}

We now setup our dynamic activity model. Formally, we denote by
$S = (S_{1}, S_{2}, \ldots, S_{T})$ our timestamped, labeled, sequential data, where
each $S_{t} \in S,\;t = 1, 2, \ldots, T$ is a data snapshot containing all developer
activity $Y_{i}$ in the $t$-time slice. Each $S_{t}$ is arranged chronologically, and the
length of the slice is a multiple of a standard time unit (e.g., weeks, months).

We adapted the Skipgram with negative sampling~\cite{mikolov2013distributed} (SGNS)
model to construct the embeddings. This method represents each developer activity
$Y_{i}$ in $S_{t}$ by two low dimensional vectors: a developer activity vector
$\bm{\mathcal{Y}}_{i}$ and a context vector $\bm{c}_{i}$.  SGNS optimizes the vectors
via stochastic gradient descent so that
$\hat{p}\left( c_{i}\,\middle\vert\,Y_{i} \right) \propto \text{exp}(\bm{\mathcal{Y}}_{i} \cdot \bm{c}_{i})$,
where $\hat{p}\left( c_{i}\,\middle\vert\,Y_{i} \right)$ is the empirical probability of
seeing $c_{i}$ within a fixed-length window of developer activities in $S_{t}$, given
that it contains $Y_{i}$. We express this window
$\mathcal{N}_{m}(Y_{i}) = \{Y_{j} \mid \abs{\Delta t(Y_{i}, Y_{j})} \leq m, Y_{j} \in Y\}$,
where $m$ is the time interval, and $Y$ is the set of all unique activities in $S$. The
size of $\mathcal{N}_{m}(Y_{i})$ and thus the number of nearby developer activities to
be considered by SGNS depends upon the value of $m$. For example,
$\mathcal{N}_{4\text{hr}}(Y_{i})$ would return all the developer activities $4$ hours
away from $Y_{i}$ across the kernel's subsystems. This is inspired by Shen and
Stringhini's sliding window approach~\cite{shenattack2vec2019}. Our model
operationalizes the ``distributional hypothesis’' that developer activities occurring in
the same context tend to have similar meanings,  and assumes that some
of them even exhibit inter-activity dependencies --- e.g., they might follow, block, or depend on one another ~\cite{maalej2017using}.

We trained our model on the entire data $S$. We followed the recommendations of Levy et
al.~\cite{levy2015improving} in setting the hyperparameters for SGNS, but used
stochastic hyperparameter optimization to set key settings. Also, we learned embeddings
of size $120$, and chose \textit{weeks} as the length of the $t$-time slices.

\subsection{Aligning Dynamic Developer Activity Embeddings}%
\label{subsec:alignment}

\begin{figure*}[t]
  \centering
  \includegraphics[width=.8\textwidth]{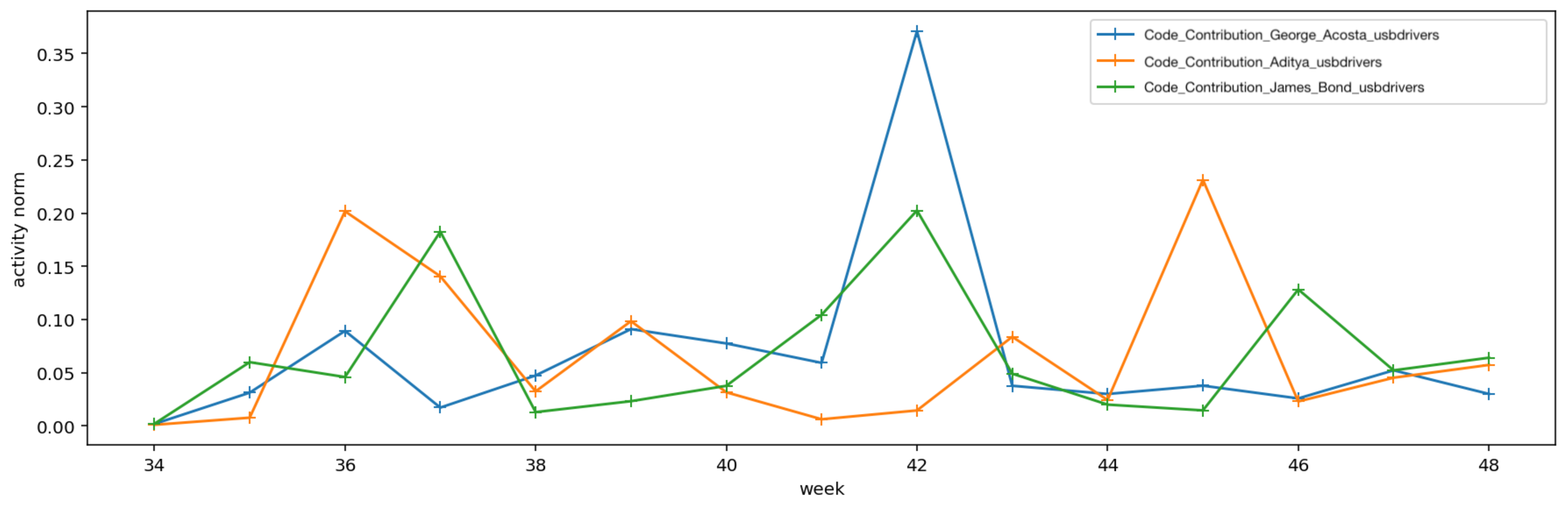}
  \caption{Evolution of contribution activities by individuals likely involved in the
    Hypocrite Commits Incident}%
  \label{fig:evolution_hypo_commits}
\end{figure*}

Charting the trajectory of influence pathways across time requires the
\textit{alignment} of the $\bm{\mathcal{Y}}_{i}$ vectors to the same coordinate axes
(aka, latent space). The low-dimensional embeddings we generated previously are
naturally unaligned due to the stochastic nature of SGNS.\@ Particularly, the training
of SGNS  could result in arbitrary orthogonal transformations that prevent the
comparison of the same developer activity across time~\cite{hamilton2016diachronic}.
Like Hamilton et al.~\cite{hamilton2016diachronic}, we first rely on training the
embeddings separately on the $S_{t}$, then use orthogonal Procrustes between every two
adjacent $t$-time slices to align the learned low-dimensional embeddings
$\bm{\mathcal{Y}}_{i}$. Specifically, we define a matrix of embeddings
$\bm{\mathscr{Y}}^{(t)} \in \mathbb{R}^{T\times|\bm{\mathcal{Y}}_{i}|}$, learned at time
period $t = 1, 2, \ldots, T$, then find a matrix $R^{(t)} \in \mathbb{R}^{T \times T}$
for mapping $\bm{\mathscr{Y}}^{(t)}$ to $\bm{\mathscr{Y}}^{(t + 1)}$:
$R^{(t)} = \arg\min_{Q}\lVert Q\bm{\mathscr{Y}}^{(t)} - \bm{\mathscr{Y}}^{(t+1)}\rVert_{F}$,
subject to $Q^{T}Q = I$.

The example in Figure~\ref{fig:three trajectories} shows the evolution of three
contribution activities $Y_{0}$ performed by individuals involved in the Hypocrite
Commits incident, under the aliases like \textit{George Acosta}, \textit{James Bond}
(confirmed in this full disclosure document~\cite{hypocritecommits2021}), and
\textit{Aditya}~\footnote{It is unclear whether the Aditya alias was involved in the
incident; however, we found their behavior to be similar to the behavior of the other
two.}. Each point in the trajectories is labeled with a tuple made of initialisms of
\textit{activity-developer-subsystem} and the time of occurrence; e.g., a point
representing \textit{George Acosta}'s contribution to the kernel's ``USB, driver core,
sysfs, $\ldots$'' area on week $38$ is labeled as \textit{(CCGAU, 38)}. In this
incident's examination, we evaluated the contextual activities most commonly associated
with these aliases' contribution activities at different time points. We compare their
activities' contexts and their inter-activity dependencies with maintainers' activities
to determine their level of influence. Acosta's contribution (see
Figure~\ref{fig:acosta}), for example, shifted sharply between weeks $36$ and $48$ (see
Figure~\ref{fig:evolution_hypo_commits}) in response to maintainers' \textit{monitoring}
and \textit{contribution} efforts. As the \textit{L2 norm} changes of \textit{CCGAU}
with surrounding activities increase, its context shifts became more noticeable.  Little
or no effect was observed for the activities of the other involved aliases.

We acknowledge the limitations of the two-step alignment process, described in Bamler
and Mandt~\cite{bamler2017dynamic}.\@ Despite the known limitations, the learned
embeddings successfully preserve the proximity and temporal continuity of activities, a
feature we used to identify three types of trust ascendancy operations: (1)
opportunistic (George Acosta), (2) awry (Aditya), and (3) hit or miss (James Bond). An
\textit{opportunistic} operation is achieved by individuals who rarely participate, but
when they do, they submit a flurry of emails. The other two types seemed to be moving
away from maintainers' activities, seemed partially isolated or rejected, or were
randomly executed. This is consistent with known and documented information of the incident.

\section{Future plans}%
\label{sec:future}

We position this work as an ideal influence pathway discovery, filtering and selection
method for signaling emergent trust ascendancy operations in OSS projects.

Charting trust ascendancy trajectories and their patterns with agility, accuracy, and
effectiveness at the scale of social coding platforms present substantial new scientific
and technical challenges. This section outlines how we can structure our research to
account for and address emerging issues. The basics of our future plans are thus: (1)
scaling to large volumes of socio-technical traces we can introspect to discover and
visualize influence trajectories while avoiding information overload and unnecessary and
complex, CPU-cycle-intensive data related operations; (2) capturing the relationships
and complexities of new case studies, paying close attention to the collection of a
diversity of signals through which developers' practices, tools, actions, and code can
be influenced in OSS projects; and (3) capturing the dynamism of influence operations
through their shifts in language, activity, and contexts linked downstream to source
code evolution and integrity. To resolve scalability issues, we will limit our
processing to only the most interesting activity traces (exemplars) and will use
stochastic hyperparameter optimizers to speed up learning and data mining tasks. To
resolve diversity issues, we rely on the high level of transparency of social coding
environments and their social media connections (e.g., GitHub, Stack Overflow, Twitter)
to generate a diversity of signals through which people's practices, tools, actions, and
code can be influenced. To address the dynamics issue, we use self-supervised training
in concert with Temporal Graph Networks~\cite{rossi2020tgn}, a dynamic graph processing
model to capture the motion dynamics of trust ascendancy without requiring embedding
alignments.

\bibliographystyle{IEEEtranS}
\bibliography{IEEEabrv,main}

\begin{thebibliography}{10}
\providecommand{\url}[1]{#1}
\csname url@samestyle\endcsname
\providecommand{\newblock}{\relax}
\providecommand{\bibinfo}[2]{#2}
\providecommand{\BIBentrySTDinterwordspacing}{\spaceskip=0pt\relax}
\providecommand{\BIBentryALTinterwordstretchfactor}{4}
\providecommand{\BIBentryALTinterwordspacing}{\spaceskip=\fontdimen2\font plus
\BIBentryALTinterwordstretchfactor\fontdimen3\font minus
  \fontdimen4\font\relax}
\providecommand{\BIBforeignlanguage}[2]{{%
\expandafter\ifx\csname l@#1\endcsname\relax
\typeout{** WARNING: IEEEtranS.bst: No hyphenation pattern has been}%
\typeout{** loaded for the language `#1'. Using the pattern for}%
\typeout{** the default language instead.}%
\else
\language=\csname l@#1\endcsname
\fi
#2}}
\providecommand{\BIBdecl}{\relax}
\BIBdecl

\bibitem{bamler2017dynamic}
R.~Bamler and S.~Mandt, ``Dynamic word embeddings,'' in \emph{International
  conference on Machine learning}.\hskip 1em plus 0.5em minus 0.4em\relax PMLR,
  2017, pp. 380--389.

\bibitem{bettenburg2015management}
N.~Bettenburg, A.~E. Hassan, B.~Adams, and D.~M. German, ``Management of
  community contributions,'' \emph{Empirical Software Engineering}, vol.~20,
  no.~1, pp. 252--289, 2015.

\bibitem{calefato2017preliminary}
F.~Calefato, F.~Lanubile, and N.~Novielli, ``A preliminary analysis on the
  effects of propensity to trust in distributed software development,'' in
  \emph{2017 IEEE 12th international conference on global software engineering
  (ICGSE)}.\hskip 1em plus 0.5em minus 0.4em\relax IEEE, 2017, pp. 56--60.

\bibitem{chen2019dynamic}
C.~Chen, Y.~Tao, and H.~Lin, ``Dynamic network embeddings for network evolution
  analysis,'' \emph{arXiv preprint arXiv:1906.09860}, 2019.

\bibitem{cheng2019activity}
J.~Cheng and J.~L. Guo, ``Activity-based analysis of open source software
  contributors: Roles and dynamics,'' in \emph{2019 IEEE/ACM 12th International
  Workshop on Cooperative and Human Aspects of Software Engineering
  (CHASE)}.\hskip 1em plus 0.5em minus 0.4em\relax IEEE, 2019, pp. 11--18.

\bibitem{child2006essentials}
D.~Child, \emph{The essentials of factor analysis}.\hskip 1em plus 0.5em minus
  0.4em\relax A\&C Black, 2006.

\bibitem{lkml_Cook_2021}
\BIBentryALTinterwordspacing
K.~Cook, ``{Report on University of Minnesota Breach-of-Trust Incident},'' May
  2021, accessed: 2022-05-01. [Online]. Available:
  \url{https://lkml.org/lkml/2021/5/5/1244}
\BIBentrySTDinterwordspacing

\bibitem{csuvik2019source}
V.~Csuvik, A.~Kicsi, and L.~Vid{\'a}cs, ``Source code level word embeddings in
  aiding semantic test-to-code traceability,'' in \emph{2019 IEEE/ACM 10th
  International Symposium on Software and Systems Traceability (SST)}.\hskip
  1em plus 0.5em minus 0.4em\relax IEEE, 2019, pp. 29--36.

\bibitem{dabbish2012social}
L.~Dabbish, C.~Stuart, J.~Tsay, and J.~Herbsleb, ``Social coding in github:
  transparency and collaboration in an open software repository,'' in
  \emph{Proceedings of the ACM 2012 conference on computer supported
  cooperative work}, 2012, pp. 1277--1286.

\bibitem{dakhel2022dev2vec}
A.~M. Dakhel, M.~C. Desmarais, and F.~Khomh, ``Dev2vec: Representing domain
  expertise of developers in an embedding space,'' \emph{arXiv preprint
  arXiv:2207.05132}, 2022.

\bibitem{de2010can}
P.~B. De~Laat, ``How can contributors to open-source communities be trusted? on
  the assumption, inference, and substitution of trust,'' \emph{Ethics and
  information technology}, vol.~12, no.~4, pp. 327--341, 2010.

\bibitem{di2019training}
V.~Di~Carlo, F.~Bianchi, and M.~Palmonari, ``Training temporal word embeddings
  with a compass,'' in \emph{Proceedings of the AAAI conference on artificial
  intelligence}, vol.~33, no.~01, 2019, pp. 6326--6334.

\bibitem{ducheneaut2005socialization}
N.~Ducheneaut, ``Socialization in an open source software community: A
  socio-technical analysis,'' \emph{Computer Supported Cooperative Work
  (CSCW)}, vol.~14, no.~4, pp. 323--368, 2005.

\bibitem{filippova2016effects}
A.~Filippova and H.~Cho, ``The effects and antecedents of conflict in free and
  open source software development,'' in \emph{Proceedings of the 19th ACM
  Conference on Computer-Supported Cooperative Work \& Social Computing}, 2016,
  pp. 705--716.

\bibitem{gysin2010trustability}
F.~S. Gysin and A.~Kuhn, ``A trustability metric for code search based on
  developer karma,'' in \emph{Proceedings of 2010 icse workshop on
  search-driven development: Users, infrastructure, tools and evaluation},
  2010, pp. 41--44.

\bibitem{hamilton2016diachronic}
W.~L. Hamilton, J.~Leskovec, and D.~Jurafsky, ``Diachronic word embeddings
  reveal statistical laws of semantic change,'' \emph{arXiv preprint
  arXiv:1605.09096}, 2016.

\bibitem{hofmann-etal-2021-dynamic}
\BIBentryALTinterwordspacing
V.~Hofmann, J.~Pierrehumbert, and H.~Sch{\"u}tze, ``Dynamic contextualized word
  embeddings,'' in \emph{Proceedings of the 59th Annual Meeting of the
  Association for Computational Linguistics and the 11th International Joint
  Conference on Natural Language Processing (Volume 1: Long Papers)}.\hskip 1em
  plus 0.5em minus 0.4em\relax Online: Association for Computational
  Linguistics, Aug. 2021, pp. 6970--6984. [Online]. Available:
  \url{https://aclanthology.org/2021.acl-long.542}
\BIBentrySTDinterwordspacing

\bibitem{jiang2014tracing}
Y.~Jiang, B.~Adams, F.~Khomh, and D.~M. German, ``Tracing back the history of
  commits in low-tech reviewing environments: a case study of the linux
  kernel,'' in \emph{Proceedings of the 8th ACM/IEEE International Symposium on
  Empirical Software Engineering and Measurement}, 2014, pp. 1--10.

\bibitem{kroahhartman2007linux}
G.~KroahHartman \emph{et~al.}, ``Linux kernel development,'' in \emph{Linux
  Symposium}.\hskip 1em plus 0.5em minus 0.4em\relax Citeseer, 2007, pp.
  239--244.

\bibitem{kulkarni2015statistically}
V.~Kulkarni, R.~Al-Rfou, B.~Perozzi, and S.~Skiena, ``Statistically significant
  detection of linguistic change,'' in \emph{Proceedings of the 24th
  international conference on world wide web}, 2015, pp. 625--635.

\bibitem{levy2015improving}
O.~Levy, Y.~Goldberg, and I.~Dagan, ``Improving distributional similarity with
  lessons learned from word embeddings,'' \emph{Transactions of the association
  for computational linguistics}, vol.~3, pp. 211--225, 2015.

\bibitem{hypocritecommits2021}
\BIBentryALTinterwordspacing
K.~Lu, ``A full disclosure of the case study of the hypocrite commits,'' 2021,
  accessed: 2022-09-01. [Online]. Available:
  \url{https://www-users.cse.umn.edu/~kjlu/papers/full-disclosure.pdf}
\BIBentrySTDinterwordspacing

\bibitem{maalej2017using}
W.~Maalej, M.~Ellmann, and R.~Robbes, ``Using contexts similarity to predict
  relationships between tasks,'' \emph{Journal of Systems and Software}, vol.
  128, pp. 267--284, 2017.

\bibitem{mahdavi2018dynnode2vec}
S.~Mahdavi, S.~Khoshraftar, and A.~An, ``dynnode2vec: Scalable dynamic network
  embedding,'' in \emph{2018 IEEE international conference on big data (Big
  Data)}.\hskip 1em plus 0.5em minus 0.4em\relax IEEE, 2018, pp. 3762--3765.

\bibitem{mikolov2013distributed}
T.~Mikolov, I.~Sutskever, K.~Chen, G.~S. Corrado, and J.~Dean, ``Distributed
  representations of words and phrases and their compositionality,''
  \emph{Advances in neural information processing systems}, vol.~26, 2013.

\bibitem{nagle2020report}
F.~Nagle, D.~Wheeler, H.~Lifshitz-Assaf, H.~Ham, and J.~Hoffman, ``Report on
  the 2020 foss contributor survey,'' \emph{The Linux Foundation Core
  Infrastructure Initiative}, 2020.

\bibitem{riehle2019analysis}
D.~Riehle, ``Analysis of ignored patches in the linux kernel development,''
  Ph.D. dissertation, Friedrich-Alexander-Universit{\"a}t
  Erlangen-N{\"u}rnberg, 2019.

\bibitem{rossi2020tgn}
\BIBentryALTinterwordspacing
E.~Rossi, B.~Chamberlain, F.~Frasca, D.~Eynard, F.~Monti, and M.~M. Bronstein,
  ``Temporal graph networks for deep learning on dynamic graphs,'' \emph{CoRR},
  vol. abs/2006.10637, 2020. [Online]. Available:
  \url{https://arxiv.org/abs/2006.10637}
\BIBentrySTDinterwordspacing

\bibitem{rudolph2018dynamic}
M.~Rudolph and D.~Blei, ``Dynamic embeddings for language evolution,'' in
  \emph{Proceedings of the 2018 world wide web conference}, 2018, pp.
  1003--1011.

\bibitem{scacchi2007free}
W.~Scacchi, ``Free/open source software development,'' in \emph{Proceedings of
  the the 6th joint meeting of the European software engineering conference and
  the ACM SIGSOFT symposium on The foundations of software engineering}, 2007,
  pp. 459--468.

\bibitem{schneider2016differentiating}
D.~Schneider, S.~Spurlock, and M.~Squire, ``Differentiating communication
  styles of leaders on the linux kernel mailing list,'' in \emph{Proceedings of
  the 12th International Symposium on Open Collaboration}, 2016, pp. 1--10.

\bibitem{segev2016learn}
N.~Segev, M.~Harel, S.~Mannor, K.~Crammer, and R.~El-Yaniv, ``Learn on source,
  refine on target: A model transfer learning framework with random forests,''
  \emph{IEEE transactions on pattern analysis and machine intelligence},
  vol.~39, no.~9, pp. 1811--1824, 2016.

\bibitem{shenattack2vec2019}
\BIBentryALTinterwordspacing
Y.~Shen and G.~Stringhini, ``{ATTACK2VEC}: Leveraging temporal word embeddings
  to understand the evolution of cyberattacks,'' in \emph{28th USENIX Security
  Symposium (USENIX Security 19)}.\hskip 1em plus 0.5em minus 0.4em\relax Santa
  Clara, CA: USENIX Association, Aug. 2019, pp. 905--921. [Online]. Available:
  \url{https://www.usenix.org/conference/usenixsecurity19/presentation/shen}
\BIBentrySTDinterwordspacing

\bibitem{sinha2011entering}
V.~S. Sinha, S.~Mani, and S.~Sinha, ``Entering the circle of trust: developer
  initiation as committers in open-source projects,'' in \emph{Proceedings of
  the 8th Working Conference on Mining Software Repositories}, 2011, pp.
  133--142.

\bibitem{von2003community}
G.~Von~Krogh, S.~Spaeth, and K.~R. Lakhani, ``Community, joining, and
  specialization in open source software innovation: a case study,''
  \emph{Research policy}, vol.~32, no.~7, pp. 1217--1241, 2003.

\bibitem{wang2019persuasion}
X.~Wang, W.~Shi, R.~Kim, Y.~Oh, S.~Yang, J.~Zhang, and Z.~Yu, ``Persuasion for
  good: Towards a personalized persuasive dialogue system for social good,''
  \emph{arXiv preprint arXiv:1906.06725}, 2019.

\bibitem{wang2016diffusion}
Y.~Wang and D.~Redmiles, ``The diffusion of trust and cooperation in teams with
  individuals' variations on baseline trust,'' in \emph{Proceedings of the 19th
  ACM Conference on Computer-Supported Cooperative Work \& Social Computing},
  2016, pp. 303--318.

\bibitem{wermke2022committed}
D.~Wermke, N.~W{\"o}hler, J.~H. Klemmer, M.~Fourn{\'e}, Y.~Acar, and S.~Fahl,
  ``Committed to trust: A qualitative study on security \& trust in open source
  software projects,'' in \emph{Proceedings of the 43rd IEEE Symposium on
  Security and Privacy, IEEE S\&P 2022, May 22-26, 2022}.\hskip 1em plus 0.5em
  minus 0.4em\relax IEEE Computer Society, May 2022.

\bibitem{wu2021feasibility}
Q.~Wu and K.~Lu, ``On the feasibility of stealthily introducing vulnerabilities
  in open-source software via hypocrite commits,'' \emph{Proc. Oakland, page to
  appear}, 2021.

\bibitem{xu2018multi}
Y.~Xu and M.~Zhou, ``A multi-level dataset of linux kernel patchwork,'' in
  \emph{2018 IEEE/ACM 15th International Conference on Mining Software
  Repositories (MSR)}.\hskip 1em plus 0.5em minus 0.4em\relax IEEE, 2018, pp.
  54--57.

\bibitem{yao2018dynamic}
Z.~Yao, Y.~Sun, W.~Ding, N.~Rao, and H.~Xiong, ``Dynamic word embeddings for
  evolving semantic discovery,'' in \emph{Proceedings of the eleventh acm
  international conference on web search and data mining}, 2018, pp. 673--681.

\end{thebibliography}

\end{document}